# All-group IV transferable membrane for room-temperature mid-infrared photodetectors

Mahmoud R. M. Atalla,[1,‡] Simone Assali,[1,‡] Anis Attiaoui,[1] Cédric Lemieux-Leduc,[1] Aashish Kumar,[1] Salim Abdi,[1] and Oussama Moutanabbir[1,*]

[1]*Department of Engineering Physics, École Polytechnique de Montréal, C. P. 6079, Succ. Centre-Ville, Montréal, Québec H3C 3A7, Canada*

**ABSTRACT:** Semiconductor membranes emerged as a versatile class of nanomaterials to control lattice strain and engineer complex heterostructures enabling a variety of innovative applications. With this perspective, herein we exploit this platform to tune simultaneously the lattice parameter and bandgap energy in group IV GeSn semiconductor alloys. As Sn content is increased to reach a direct bandgap, these semiconductors become metastable and typically compressively strained. We show that the relaxation in released membranes extends the absorption wavelength range deeper in the mid-infrared. Fully released $Ge_{0.83}Sn_{0.17}$ membranes were integrated on silicon and used in the fabrication of broadband photodetectors operating at room temperature with a record wavelength cutoff of 4.6 μm, without compromising the performance at shorter wavelengths down to 2.3 μm. These membrane devices are characterized by two orders of magnitude reduction in dark current as compared to as-grown strained epitaxial layers. A variety of experimental tools and optimized calculations were used to discuss the crystalline quality, composition uniformity, lattice strain, and the electronic band structure of the investigated materials and devices. The ability to engineer all-group IV transferable mid-infrared photodetectors lays the groundwork to implement scalable and flexible sensing and imaging technologies exploiting these integrative, silicon-compatible strained-relaxed GeSn membranes.

**Introduction**

Strain engineering has been a ubiquitous paradigm to tailor the electronic band structure and harness the associated new or enhanced fundamental properties in semiconductors. In this regard, semiconductor membranes are considered as a rich platform for lattice engineering allowing precise control of the electrical, thermal, and optical properties.[1] They also enable new degrees of freedom to implement heterostructures and devices that are hardly achievable by conventional heteroepitaxial or wafer bonding processes. Indeed, three-dimensional complex membrane structures can be realized by selectively releasing thin semiconductor layers, including low-dimensional structures, tubes, micro/nano-wires, ribbons, open meshes, and flexible devices.[2] Moreover, membranes have also been a playground to uncover, investigate, and harness new fundamental properties in semiconductors enabling the demonstration of novel phenomena and processes ranging from quantum photonic interfaces[3] to thermomechanical stabilities,[4,5] in addition to their use as strain-engineered compliant substrates for lattice-mismatched epitaxial growth.[6] One of the key features is that thin membranes can be easily transferred and integrated onto flexible, curved, or transparent substrates of virtually any material, unlocking a range of new applications in inorganic flexible, stretchable, and wearable devices.[7] As a matter of fact, semiconductor membranes are currently used as versatile building blocks in a variety of applications in thermo-photovoltaics,[8] single-cell sensing,[9] pixellated solar cells,[10] light-emitting diodes,[11] photodetectors,[12,13] and lasers.[14] The large-scale integration using rolling-based direct-transfer printing method has paved the way to mass produce these high-performance flexible and hybrid devices.[15]

Broadband membrane photodetectors (PDs) are among the successfully implemented devices achieved using multiple materials including Si,[16,17] Ge,[18–20] SiGe,[21] InP,[22] InGaAs,[23] and Sb-based superlattices.[10,24] However, while these devices mainly operate at near infrared (NIR) and short-wave infrared (SWIR) wavelengths or at a longer THz wavelengths, developing membrane PDs operating in the mid-infrared (MIR) range has been severely limited by the lack of suitable material systems.[24] The ability to develop transferable MIR PDs using semiconductor released membranes is highly attractive to exploit all attributes of this platform in addition to circumventing the high-cost and limited integration of the current technologies that are based on InSb, PbSe, and HgCdTe semiconductors.[25–33] To simultaneously address these challenges, herein we developed all-group IV membranes using strain-engineered GeSn, integrated them in the fabrication of PDs, and demonstrated their operation at room-temperature in the MIR range.

GeSn semiconductors have attracted a great deal of interest because of their Si-compatibility and the flexibility they offer to engineer lattice parameter and bandgap energy and directness. GeSn direct bandgap is achieved at a Sn content around 10 at.% in fully relaxed layers, leading to the recent demonstration of room temperature optical emission up to ~4 μm,[34–36] and optically-pumped lasers at 3.1-3.4 μm operating at temperatures of 180-270 K.[37–39]. PDs were also demonstrated using GeSn layers showing a room temperature performance close to that of commercial PbSe detectors at wavelengths reaching 3 μm,[40–43] although the residual strain limits the operation wavelength range. Indeed, GeSn semiconductors are inherently metastable and typically exhibit a compositional gradient and a large compressive strain resulting from the canonical, lattice-mismatched growth on Ge.[44,45] This compressive strain increases the Sn critical content to reach the indirect-to-direct bandgap transition thus exacerbating the growth hurdles. Moreover, at a fixed Sn content the corresponding optical emission and detection range of compressively strained layers shifts to shorter wavelengths in comparison to fully relaxed layers.[46–48] Therefore, developing GeSn membranes will not only add a new dimension in device processing, but it will also provide a path to engineer the fundamental properties of GeSn. In this regard, recent attempts focused on amorphous GeSn[49] and rolled-up GeSn/Ge membranes.[50] However, their demonstrated performance was limited to the visible-NIR range due to the low Sn content.

In this work, we show tunable GeSn photodetectors operating at room-temperature in the 1.3 to 4.6 µm wavelength range by controlling lattice strain and achieving a uniform Sn content up to 17 at.%. By releasing $Ge_{0.83}Sn_{0.17}$ membranes and transferring them onto $SiO_2$/Si substrates, the residual compressive strain is significantly relaxed, and the room-temperature spectral response of the membrane PD is further extended to 4.6 µm from 3.5 µm in as-grown layers. The redshift of the cutoff wavelength by more than 1 µm in the MIR range does not compromise the performance at shorter wavelengths down to 2.3 µm. A strong reduction in dark current was observed in membrane devices, which is critical to cut down the dissipated power and the total device noise and enable the operation at higher bias voltage as compared to conventional devices fabricated using as-grown epitaxial layers.

**Results and Discussion**

**Growth and characterization of GeSn epilayers.** The epitaxial growth protocols were first optimized to achieve different GeSn multilayers with uniform Sn contents and variable strain levels. The growth was carried out on Si wafers using Ge as an interlayer (Ge virtual substrate, Ge-VS) in a low-pressure chemical vapor deposition (CVD) reactor (see Methods for details). The investigated sets of samples are referred to by the composition of the topmost GeSn layer corresponding to the device active layer, namely $Ge_{0.895}Sn_{0.105}$, $Ge_{0.87}Sn_{0.13}$, and $Ge_{0.83}Sn_{0.17}$, which have thicknesses of 150, 120, and 160 nm, respectively. Their structural parameters are listed in Supporting Information S1. The cross-sectional transmission electron microscopy (TEM) image and electron energy loss spectroscopy (EELS) profile of the $Ge_{0.87}Sn_{0.13}$ sample are shown in Fig. 1a-b. The multi-layered growth promotes strain relaxation and ensures an enhanced uniform incorporation of Sn. The composition of each layer is estimated from Reciprocal Space Mapping (RSM) around the asymmetrical (224) X-ray diffraction (XRD) peak (Fig.1c-e). For $Ge_{0.87}Sn_{0.13}$, four GeSn layers with Sn contents of 5.0 at.% (#1), 6.3 at.% (#2), 8.8 at.% (#3) and 11.9 at.% (#4) were used as buffer layers to grow the GeSn top layer (TL) with high crystalline quality and a uniform Sn content of 13.3 at.%. A total GeSn thickness of ~1100 nm-thick was grown on a 1.6 µm-thick Ge-VS. Gliding of misfit dislocations at the interface between GeSn layers with different Sn content took place during the growth, while threading dislocations are not detected at the TEM imaging scale in the upper layers, confirming the high crystalline quality of the GeSn TL. For $Ge_{0.895}Sn_{0.105}$, three GeSn buffer layers were used. A high degree of strain relaxation is obtained for both $Ge_{0.895}Sn_{0.105}$ (Fig. 1c) and $Ge_{0.87}Sn_{0.13}$ (Fig. 1d) leading to a residual compressive in-plane biaxial strain $\varepsilon_{||}$ below $-0.5$ % in the TL, without any compositional grading. Note that only two buffer layers $Ge_{0.92}Sn_{0.08}$ (#1) and $Ge_{0.88}Sn_{0.12}$ (#2) were sufficient to optimize the growth at higher Sn content in $Ge_{0.83}Sn_{0.17}$ (TL), as shown in the RSM data (Fig. 1e). A larger residual compressive strain $\varepsilon_{||} \sim -1.3$ % is, however, measured in this set of samples (Fig. 1e) without compromising the crystalline quality and composition uniformity.[35,45]

Room temperature photoluminescence (PL) recorded from the as-grown layers, before any device fabrication, are displayed in Fig. 1f. A clear single emission peak is visible at 2.4 µm in $Ge_{0.895}Sn_{0.105}$, while the PL peak shifts to 3.0 µm and 3.4 µm in $Ge_{0.87}Sn_{0.13}$ and $Ge_{0.83}Sn_{0.17}$, respectively. The PL emission at room temperature hints to the bandgap directness of these layers, in good agreement with theoretical calculations for GeSn alloy with Sn content higher than 10 at.% [51]. These measurements confirm that the selected compositional range yields bandgap energies in the MIR range. Moreover, the observed clear room temperature PL emission is also an indication of their high crystalline quality of the samples. It is noteworthy that the broader PL in $Ge_{0.87}Sn_{0.13}$ emission results from the contribution of the 11.9 at.% buffer layer (#4) to the emission due to the small band offset (< 30 meV) with the 13.3 at.% TL.[36] In contrast, the PL emission of $Ge_{0.83}Sn_{0.17}$ originates mainly from the 17 at.% layer because of the larger band offset (40-60 meV) with the 12 at.% layer underneath.[35,36] It is important to note that the compressive strain increases the band gap energy of GeSn as compared to a relaxed material at a fixed composition.[36,46–48] Therefore, these strained layers can greatly benefit from the membrane release to relax the residual strain beside the additional flexibility in device processing and performance, as discussed below.

**GeSn membrane release and transfer.** In the following, the focus will be on the highly strained, high Sn content $Ge_{0.83}Sn_{0.17}$ layers. The latter were released from the Ge/Si substrates by patterning and underetching free-standing membranes, as illustrated in Fig. 2a. First, the anisotropic reactive ion etching (RIE) of the GeSn membrane sidewalls was achieved by using a $Cl_2$-based plasma, followed by a selective etching of Ge layer using a $CF_4$-based plasma. Arrays of membrane with lateral dimensions of 10 μm × 10 μm, 10 μm × 20 μm, and 20 μm × 20 μm were fabricated, and the scanning electron microscope (SEM) image in Fig. 2b shows a single 10 μm × 20 μm membrane. The $CF_4$ etching time was selected to completely release the membranes until they collapse on Si wafers, resulting in membranes with a thickness of ~320 nm corresponding to the thickness of $Ge_{0.83}Sn_{0.17}/Ge_{0.88}Sn_{0.12}$ layers (Fig. 2b), indicating a complete underetching of Ge and Ge$_{0.92}$Sn$_{0.08}$ buffer layer. Note that the released membranes exhibit a downward bow, which is expected due to the unbalanced strain in the released Ge$_{0.83}$Sn$_{0.17}$/Ge$_{0.88}$Sn$_{0.12}$ bilayer membrane.[52] To evaluate the residual strain in the released membranes, confocal Raman spectroscopy measurements were performed. The 2D Raman maps recorded at the Ge-Ge longitudinal optical (LO) phonon peak position for as-grown and membrane samples are displayed in Fig. 2c-e. A representative set of local Raman spectra are shown in Supporting Information S2. From these maps, the Ge-Ge mode is observed at 292.5 ± 1.0 cm$^{-1}$ in the as-grown layer (Fig. 2c), but shifts to 287.0 ±1.0 cm$^{-1}$ at the edge of the partially released membrane (Fig. 2d), and to 285.0 ± 1.0 cm$^{-1}$ in a completely released membrane (Fig. 2e). It is noticeable that the Ge-Ge peak position varies slightly across the completely released membrane plausibly due to local fluctuations in strain. Note that Raman spectra associated to the area between membranes only show the Si-Si mode at 520 cm$^{-1}$ originating from the substrate, thus confirming that the Ge layer is completely etched. A close examination of Raman spectra (see supporting information S2) shows that the Ge-Sn peak also shifts from 255 cm$^{-1}$ to 246 cm$^{-1}$ in GeSn membranes. However, the full width half maximum does not increase for both Ge-Ge and Ge-Sn peaks upon strain relaxation. Based on recent systematic studies decoupling Sn content and lattice strain and allows to calculate the lattice strain since the Sn content is known,[53] we confirm from Raman data a change in strain from the initial -1.3% to roughly ±0.2% in the released membranes. This significant relaxation and the local residual strain fluctuations are expected to affect the optical properties of GeSn membranes and thus the performance of PDs.

**Band structure calculations.** Before delving into the device properties, we first discuss the band structure of GeSn membranes. A theoretical framework is developed to allow a quantitative description of the band structure of GeSn as a function of strain and composition exploiting the 8-band $k.p$ model with envelope function approximation (EFA)[54]. To account for the bowing effects in the L and Γ Brillouin zone directions, we carried out a thorough PL study of the band gap energy temperature dependence.[36] This allows for a new set of bowing parameters to be evaluated and fitted to the 8-band $k.p$ energy. The bowing parameters (at 300 K) $b_\Gamma = 2.18$ eV and $b_L = 0.68$ eV were extracted and used, and the Bir-Pikus formalism [55] was introduced to account for the lattice strain. The Ge-Ge LO phonon mode map of a transferred and processed GeSn membrane (Fig. 3a) is displayed in Fig. 3b. The residual biaxial strain after device processing is then evaluated following the procedure described in the Supporting Information S2. Fig.3c exhibits the corresponding strain map indicating a strong relaxation with local strain taking values between -0.2 and +0.2%, confirming the observations outlined in Fig. 2. From this strain map, the energy band gap map of the $Ge_{0.83}Sn_{0.17}$ membrane along the Γ Brillouin direction is calculated with the 8-band $k.p$ formalism, as shown in Fig. 3d. Overall the band gap energy has expectedly shrunk upon relaxation from 0.36 eV in the as-grown layer to 0.25-0.31 eV in the released membrane. In Fig. 3e, the calculated band lineup (at 300 K) is outlined for the as-grown $Ge/Ge_{0.92}Sn_{0.08}/Ge_{0.88}Sn_{0.12}/Ge_{0.83}Sn_{0.17}$ stack (full-line in Fig. 3e) and compared to $Ge_{0.83}Sn_{0.17}$ membrane at two different residual strain values -0.2% (dashed-line) and +0.2% (dotted-line). The valence band and conduction band offsets change from 64 meV and 18 meV in as-grown $Ge_{0.88}Sn_{0.12}/Ge_{0.83}Sn_{0.17}$ layers to ~40 meV and ~90 meV upon strain relaxation in the released membrane, respectively. From the calculated band energy diagram, it is clear that both holes and electrons will diffuse to $Ge_{0.83}Sn_{0.17}$ TL even at room temperature owing to the large band offsets.

**Fabrication and characterization of as-grown GeSn photodetectors.** To discuss the performance of the GeSn membrane PDs, it is important to establish first the baseline knowledge of the behavior of GeSn layers immediately after the growth. To this end, photoconductive devices were fabricated using the as-grown GeSn layers at different Sn content. These devices have an interdigitated geometry and two Ti/Au (10 nm/60 nm) metal contacts. A schematic and an optical micrograph of a typical GeSn interdigitated PD are shown in Fig. 4a-b. The dark current as a function of the applied bias for all samples is shown in Fig. 4c, indicating a steep increase as Sn content increases, eventually reaching a factor ~8 increase in $Ge_{0.83}Sn_{0.17}$. The higher leakage current in GeSn is believed to originate from the presence of point defects, such as vacancies and vacancy complexes,[53] which have been suggested to be responsible for the unintentional p-type doping in GeSn that is an order of magnitude higher than in epitaxial Ge.[56–58] It is worth mentioning that other factors could play a role in the measured dark current, such as the absence of a surface passivation layer[43] and the parallel conductance associated with the Ge-VS/Si stacking.[59] Since the incident light has an achromatic spot size of ~7 μm at the sample after the reflective objective, the photocurrent magnitude is thus independent of the larger active membrane device area. The spectral responsivity for all samples acquired at room temperature with an electric field of 0.16 kV/cm are shown in Fig. 4d. The highest responsivity is achieved in Ge, reaching up to ~0.1 A/W at a wavelength λ = 1.5 μm, close to the device cutoff at 1.75 μm. Note that a responsivity of ~1 A/W is obtained at an electric field of 1.6 kV/cm, which emphasizes the expected linear behavior of the responsivity of Ge PDs *vs.* the electric field. For GeSn PDs, the cutoff shifts to longer wavelengths as Sn content increases, reaching 2.6 μm, 3.15 μm, and 3.5 μm in $Ge_{0.895}Sn_{0.105}$, $Ge_{0.87}Sn_{0.13}$, and $Ge_{0.83}Sn_{0.17}$, respectively. This shift is in agreement with the PL measurements exhibited in Fig. 1f. We note that a lower responsivity (<40 mA/W) is measured in GeSn as compared to Ge at a wavelength of 1.5 µm. The high density of dislocations in the bottom region of the GeSn multi-layer stacking (Fig. 1a) most likely suppresses the collection of the optically generated carriers.

As we move toward longer wavelengths, a monotonic decrease in the responsivity of GeSn PDs is observed, starting from ~4 mA/W at 1.5 µm ($Ge_{0.895}Sn_{0.105}$) and reaching ~0.2 mA/W at 2.3 µm ($Ge_{0.895}Sn_{0.105}$), ~0.1 mA/W at 2.8 µm ($Ge_{0.87}Sn_{0.13}$), and ~0.02 mA/W at 3.2 µm ($Ge_{0.83}Sn_{0.17}$), close to the device cutoff (Fig. 4d). The decrease in responsivity toward longer wavelengths originates from the reduced effective thickness of GeSn layers that can absorb the incoming MIR radiation. By increasing the Sn content, the band gap shrinks (Fig. 3), and the absorption edge extends to longer wavelengths, hence only the narrower bandgap in the topmost layers of the stacking still contributes to absorption. For example, photocurrent signal at wavelengths beyond 3.0 µm would only be associated to the optical absorption in $Ge_{0.83}Sn_{0.17}$, while at shorter wavelengths the underlying lower Sn content layers also contribute to the absorption thus enhancing the PD responsivity. Trapping centers in GeSn are mainly located in the defective low Sn content layers,[45] thus they are expected to play a minor role in the PD performance. As an indication of this, similar responsivities were obtained at a fixed detection wavelength in GeSn PDs with different layer thickness and composition (see 2.0-2.5 µm range in Fig. 4d).

The dependence of the $Ge_{0.83}Sn_{0.17}$ PD responsivity on the applied electric field is displayed in Fig. 4e. Since no qualitative changes in the spectral responsivity with the applied electric field were observed, only a few wavelengths (1.5, 1.8, 2.4, and 3.0 µm) were selected. A consistent linear increase in responsivity with the electric field is observed across the 0.16-2.6 kV/cm range (inset in Fig. 4e), which is desirable for adjustable gain detection applications.[60] The responsivity of the $Ge_{0.83}Sn_{0.17}$ PD increases significantly as the applied electric field increases to 2.6 kV/cm, however, no photoconductive gain is measured. For more insights into the nonradiative processes, the excitation power density was varied at a fixed bias voltage. The dependence of the responsivity on the excitation power density estimated at 1.5 μm is shown in Fig. 4f. A small decrease in responsivity from 5.5 mA/W to 4.3 mA/W is measured in $Ge_{0.895}Sn_{0.105}$ PD when increasing the power density from 2 kW/cm$^2$ to 11 kW/cm$^2$, while the curve is almost flat for $Ge_{0.87}Sn_{0.13}$ and $Ge_{0.83}Sn_{0.17}$ PDs. This indicates that nonradiative recombination channels have a minimum impact on the responsivity of GeSn, in agreement with recent PL studies.[36]

From the Ohmic behavior of dark current in Ge and GeSn PDs (Figs. 4c), the specific resistivity $\rho_c$ of the device contacts and the sheet resistance $R_{sh}$ were estimated from transfer length measurements (TLM), as discussed in Supporting Information S4. The one order of magnitude reduction of $\rho_c$ observed in GeSn ($1 - 3 \times 10^{-4}\ \Omega cm^2$) compared to Ge ($2.8 \times 10^{-3}\ \Omega cm^2$) is in agreement with the higher efficiency observed for Ge as compared to GeSn. However, the doping level of the active layer plays an important role in the contact resistance. To decouple the contact and doping effects, capacitance-voltage (C-V) measurements were performed (Supporting Information S5). A p-type conductivity was obtained in the Ge reference layer with an active carrier concentration of ~ $6 \times 10^{16}$ cm$^{-3}$, while this value further increases to 1-5×10$^{17}$ cm$^{-3}$ in GeSn. This difference agrees with the estimated sheet resistance from TLM measurements, therefore indicating that the contacts are not the limiting factor in the PD performance.

**GeSn membrane photodetectors.** For device processing, the released $Ge_{0.83}Sn_{0.17}$ membranes were transferred onto SiO$_2$/Si substrates. Transferred individual membranes were processed into interdigitated PDs, as illustrated in Fig. 5a-b (see Methods for more details). The dark current as a function of the applied bias for the as-grown and membrane PDs are shown in Fig. 5c. A two orders of magnitude reduction in dark current is estimated in the membrane PD compared to the as-grown $Ge_{0.83}Sn_{0.17}$ device, as a result of the insulating substrate underneath rather than the highly conductive Ge-VS/Si stacking. In principle, the low dark current in the membrane will reduce the power dissipation and unlock device operation at higher bias voltage, while increasing the photocurrent thereby overcoming the difficult task of the current amplification in as-grown devices. Note that the I-V measurement of the membrane PD exhibits a Schottky behavior (Fig. 5c), in sharp contrast to the Ohmic behavior observed in the as-grown PD (Fig. 4c). This could be caused by the fluorine $CF_4$ plasma used to underetch Ge sacrificial layer.[61–63] Fluorine plasma effectively passivates the dangling bonds at the GeSn surface by forming Ge-F covalent bonds with partially ionic property, which decreases the number of surface states and enforces Fermi level depinning[61] leading to a p-type Schottky contacts on the membrane PD. On contrast, the unintentionally p-type doped, as-grown device has Ti/Au Ohmic contacts on $Ge_{0.83}Sn_{0.17}$ because of the Fermi level pining to the charge neutrality level which is located close to the valence band.

The spectral responsivity for $Ge_{0.83}Sn_{0.17}$ membrane and as-grown PDs are compared in Fig. 5d. Multiple features are observed in the membrane spectral response. First, a monotonous decrease in the responsivity is visible at wavelengths shorter than 2.3 µm, eventually decreasing by a factor of ~60 at 1.5 µm. Second, similar responsivity is obtained for both as-grown and membrane PDs in the 2.3-3.2 µm wavelength range. Third, at longer wavelengths the as-grown sample reaches a cutoff at 3.5 µm, in agreement with PL data (Fig. 1f), while the responsivity of the membrane extends and remains constant up to ~4.2 µm before reaching the cutoff at ~4.6 µm. This behavior demonstrates the tunability of PDs to operate across a broader MIR range by exploiting strain relaxation in GeSn membranes. It is also noticeable that the as-grown device exhibits a much abrupt cutoff (slope = $5.84 \times 10^3$ A/Wm), while the membrane device has a more gradual cutoff (slope = $1.74 \times 10^3$ A/Wm). Local fluctuations in the membrane band gap energy are most likely responsible for the reduced slope toward the cutoff. This is also in agreement with the energy band gap map in Fig. 3d showing fluctuations between 0.31 eV (4 μm) and 0.25 eV (4.9 μm). Vertical dashed lines in Fig. 5d indicate the absorption onset for Ge and GeSn bottom and middle layers with 8 and 12 at.% Sn (#1-2), respectively. The drop in the membrane PD responsivity at wavelengths below 2.3 μm is due to the under-etching of the $Ge_{0.92}Sn_{0.08}$/Ge. Indeed, absorbance measurements[36] show that the onsets for as-grown $Ge_{0.88}Sn_{0.12}$ and $Ge_{0.83}Sn_{0.17}$ are detected at 2.7 μm and 3.6 μm, respectively, in good agreement with the absorption edges observed in the responsivity curve (Fig. 4d). Thus, in the membrane PD only the $Ge_{0.88}Sn_{0.12}$ and $Ge_{0.83}Sn_{0.17}$ contribute to the photocurrent at wavelengths longer than 2.3 μm. This is further shown by the spectral responsivity ratio of the membrane w.r.t. the as-grown $Ge_{0.83}Sn_{0.17}$ ($R_{memb}/R_{as-grown}$), as displayed in Fig. 5e. Similar responsivity ratio is estimated in the 2.3-3.5 μm wavelength range, while enhanced (reduced) membrane responsivity is visible at longer (shorter) wavelengths. Thus, strain engineering in membrane extends the functionality of GeSn by covering the 3.5-4.6 μm absorption band, without compromising efficiency at shorter wavelengths down to 2.3 μm. Note that this work's focus is to establish the main experimental window of the basic material properties. However, we envision that the performance of these GeSn membrane PDs to be enhanced by surface passivation processes combined with post-growth annealing to reduce the effect of defect trapping of charge carriers. Additionally, device architecture could also be optimized to reduce the carrier transit time over carrier lifetime to enhance device responsivity via photoconductive gain. Light trapping through nanostructuring including resonant cavity, photonic band gap engineerin, and plasmonic absorption enhancement can also be explored in this system to the external quantum efficiency of the GeSn membrane devices. Moreover, the cutoff wavelength can also be extended through strain engineering using mechanical or thermal processes. For instance, 1% tensile strained GeSn membrane at 17at.% Sn content would allow PDs with a cutoff exceeding 6 μm.[64] The availability of high-quality, strain-relaxed GeSn membrane will make possible these future systematic studies.

**Conclusion**

In this work, broadband GeSn membrane photodetectors (PDs) operating at room-temperature in the MIR wavelength range are demonstrated. We showed that the PD spectral tunability in the 1.5-4.6 μm wavelength range is achieved by controlling lattice strain and Sn content in the 10.5-17 at.% range. We also established a process to release and transfer onto SiO$_2$/Si substrates of initially highly strained Ge$_{0.83}$Sn$_{0.17}$ layers. The obtained strain-relaxed membranes enabled the extension of the PD cutoff by an additional ~1 μm at room-temperature, without compromising the performance at shorter wavelengths. Additionally, two orders of magnitude reduction in dark current is observed in the membrane as compared to as-grown Ge$_{0.83}$Sn$_{0.17}$ layers. The established experimental protocol is easily extendable to a variety of host substrates, including flexible substrates and curved surfaces. These membrane devices can benefit from mainstream semiconductor processing to optimize their performance, thus creating a wealth of new opportunities to engineer new MIR applications. The latter can benefit from the high-yield and cost-effectiveness of group IV semiconductors and their compatibility with silicon technologies. These advantages are central to implement large-area, integrated MIR sensors and imagers for label-free chemical contrast in biosensing or environmental gas sensing exploiting the 3–5 µm window of transparency in Earth's atmosphere.

**Methods**

*Epitaxial growth of GeSn:* GeSn layers were grown on a 4-inch Si (100) wafer in a low-pressure chemical vapor deposition (CVD) reactor using ultra-pure $H_2$ carrier gas, 10% monogermane ($GeH_4$) and tin-tetrachloride ($SnCl_4$) precursors.[35,45,65] First, the Ge-VS was grown using a two-temperature step process at 450 and 600 °C with a thickness of 1.6 µm for the $Ge_{0.895}Sn_{0.105}$ and $Ge_{0.87}Sn_{0.13}$ samples, while in the $Ge_{0.83}Sn_{0.17}$ sample the Ge-VS was grown at 450 °C with a thickness of 0.6 µm. Next, a post-growth thermal cyclic annealing (>800 °C) process was performed to improve the crystalline quality of the Ge-VS. The growth of GeSn multi-layers was then performed at temperature below 350 °C for a variable growth time. The composition of each layer was controlled by the growth temperature,[65] while the Ge/Sn ratio in gas phase was kept constant or slightly increased with decreasing temperature to prevent phase separation during growth. In the $Ge_{0.895}Sn_{0.105}$ sample the multi-layers were grown at 340 °C (#1), 330 °C (#2), 320 °C (#3), 310 °C (TL), while in $Ge_{0.87}Sn_{0.13}$ an additional layer was grown at 300 °C (TL). In the $Ge_{0.83}Sn_{0.17}$ sample growth temperatures of 320 °C (#1), 300 °C (#2), 280 °C (TL) were used.[35,45,65]

*Raman measurements:* Raman measurement were performed using an InVia Raman microscope with a 633 nm laser and equipped with a 1800/mm diffraction grating. Note that in Raman spectra the contribution from the lower Sn content layers (#1-2) is negligible because of the penetration depth of the 633 nm excitation laser being limited to less than 30 nm, thus significantly smaller than the 17 at.% TL thickness of 160 nm. It is worth noting that the GeSn layers heating effects are eliminated by carrying out the measurements at sufficiently low laser power. Indeed, the systematic studies performed on the investigated samples indicated that a laser power density of ~3 mW/µm$^2$ does not induce any peak shift or asymmetric broadening, thus confirming the absence of laser-induced heating. As a matter of fact, lowering the laser power density even by a factor of 10 down to ~0.3 mW/µm$^2$ was found to yield identical spectra.

*Membrane photodetector fabrication:* The GeSn photodetector device fabrication started by spin coating a photoresist mask for chlorine etch using ICP Oxford Instruments PlasmaLab System100 etcher. This etching was done by flowing Cl2/N2/O2 at 40/10/4 sccm at 30 ºC, 20 mTorr pressure and 100 W RF power for 3 min. This is enough to etch all of the GeSn/Ge-VS layers reaching to the Si substrate. Following SEM inspection, the photoresist was stripped using O2 plasma asher at 400 sccm and 500 W at room temperature for 2 min. To underetch-release the GeSn layers, the Ge layer was then etched using fluorine-based etching which has high selectivity in etching Ge and preserving GeSn of rich Sn content. This fluorine etch was made by plasma asher flowing CF4 at 200 sccm and 200 W for 5 min. This fluorine releases completely the GeSn layer leaving 20 µm × 20 µm flakes that are then transferred onto semi-insulating SiO2/p++-Si substrate with 90 nm of SiO2. A transfer station from Graphene-hq company has been used to transfer the GeSn membranes onto the semi-insulating substrate which has gold grid and align marks to help aligning the contacts to these membranes in subsequent steps. Polycarbonate (PC, Sigma Aldrich, 6% dissolved in chloroform) atop PDMS was used to help picking up the membranes that have been completely underetched. The PC with the membranes is then released from the PDMS and dropped off onto the semi-insulating substrate and heated at 150 ºC to increase the contact area of PC with the SiO$_2$ and melt PC onto the substrate. The PC residue was cleaned by soaking the sample in chloroform for 10 min leaving only the GeSn membranes on top of the substrate. HCl:DI water 1:4 were used for cleaning of the transferred membranes to reduce the native oxide. This step was followed by spin coat MMA/PMMA two-layer resist for EBL patterning. The two-layer resist was made to get MMA resist thickness that is enough to lift-off a 300 nm thick metal and to get a decent undercut resist profile which helps in metal lift-off without the need for ultrasonication which could easily damage these devices. It was necessary to deposit a thick contact metal to ensure electrical connection despite the mesa formed by the GeSn thick membrane edge and its curved surface profile formed by the strain relaxation. EBL was later used to pattern the photodetector contacts on the transferred membrane aligning them in reference to the already existing gold grid on the semi-insulating substrate. The contacts were deposited using e-beam evaporation for Ti/Au (30 nm/270 nm). Lift-off were then done using remover 1165 while soaking at 70 ºC for one hour.

*Photocurrent measurements:* The I-V measurements were acquired using Keithley 4200a parameter analyzer connected to a probe station. The I-V curves of the membrane devices exhibited low dark current with Schottky behavior which most likely occurred as a result of fluorine-based etching of these flakes. The photocurrent was measured at 1.55 µm wavelength showing a nonlinear I-V curve as the bias increases. Moreover, the spectral responsivity was measured using Bruker Vertex 70 FTIR spectrometer. The MIR light source of the FTIR was incident on the GeSn device and the electrical signal was measured using a Zurich Instruments lock-in amplifier that was locked to the frequency of a chopper in the light path of the MIR light source. The lock-in signal is fed to the FTIR electronics to eventually get the photocurrent as function of wavelength. Knowing the power profile of the MIR light source the spectral responsivity of the PD can be calculated.

**Supporting Information**

The Supporting Information will be available free of charge on the Wiley Online Library or from the author.

**Author Information**

‡ These authors contributed equally to this work.

Corresponding Author:

*E-mail: oussama.moutanabbir@polymtl.ca

Notes

The authors declare no competing financial interest.

**Acknowledgements**

The authors thanks J. Bouchard for the technical support with the CVD system, B. Baloukas for support with the integrating sphere measurement. O.M. acknowledges support from NSERC Canada (Discovery, SPG, and CRD Grants), Canada Research Chairs, Canada Foundation for Innovation, Mitacs, PRIMA Québec, and Defence Canada (Innovation for Defence Excellence and Security, IDEaS).

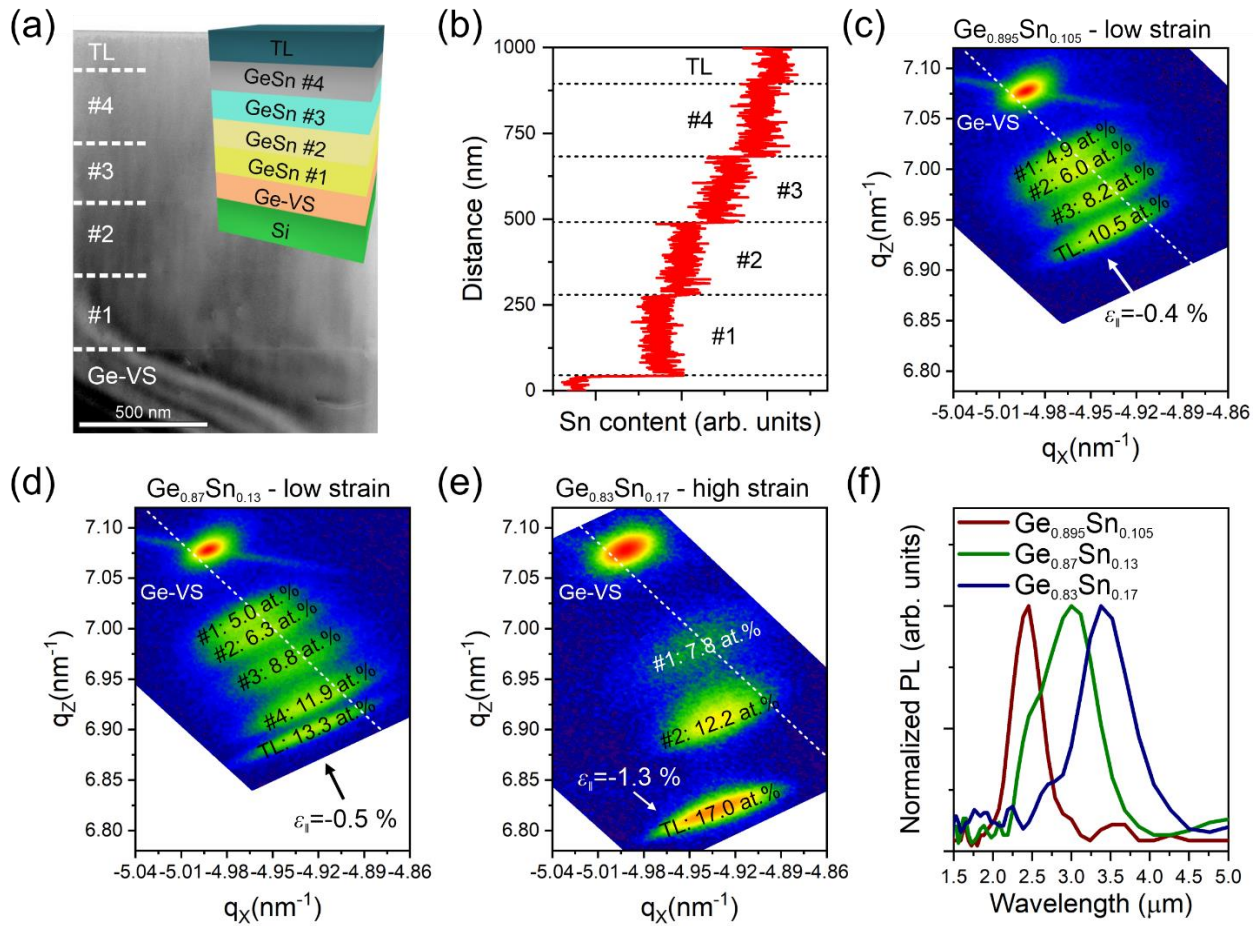

**Figure 1.** (a) Cross-sectional TEM image recorded along the [110] zone axis of as-grown Ge$_{0.87}$Sn$_{0.13}$. Inset: illustration of the multi-layer growth; (b) EELS profile of the same sample in (a); (c-e) RSM around the asymmetrical (224) reflection for the Ge$_{0.895}$Sn$_{0.105}$ (c), Ge$_{0.87}$Sn$_{0.13}$ (d), and Ge$_{0.83}$Sn$_{0.17}$ (e) layers; (f) Room-temperature PL spectra of the investigated samples.

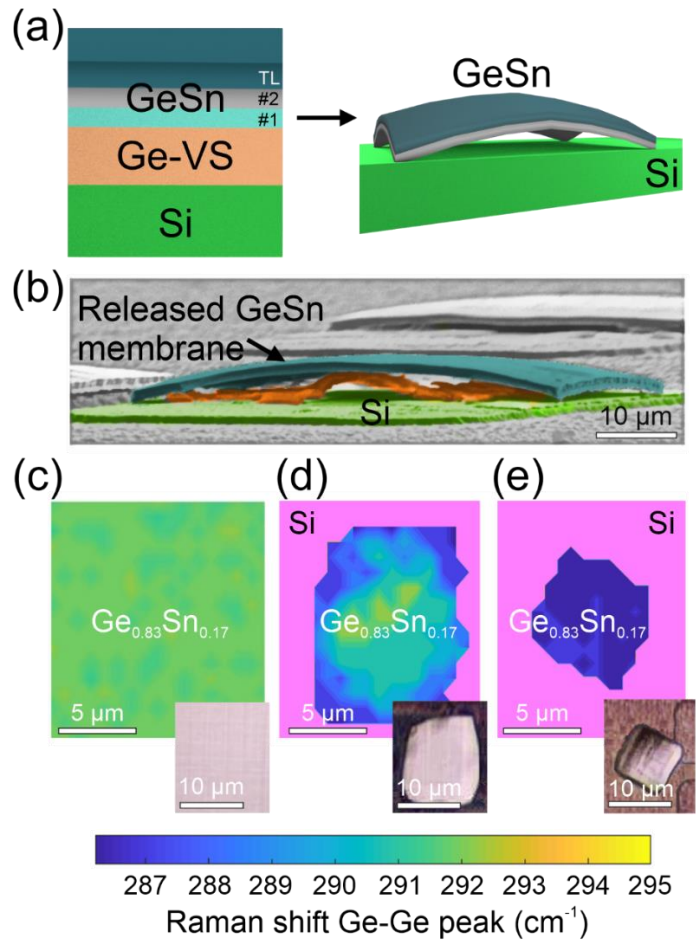

**Figure 2.** (a) Schematics of the membrane fabrication process; (b) SEM of a single released Ge$_{0.83}$Sn$_{0.17}$ membrane with the orange layer comprised Ge and low Sn content GeSn layers; (c-e) Raman spectroscopy maps recorded at the Ge-Ge LO mode for: (c) the Ge$_{0.83}$Sn$_{0.17}$ as-grown layer; (d) Partially suspended Ge$_{0.83}$Sn$_{0.17}$ membrane; and (e) fully released Ge$_{0.83}$Sn$_{0.17}$ membrane. Insets: the corresponding optical micrographs.

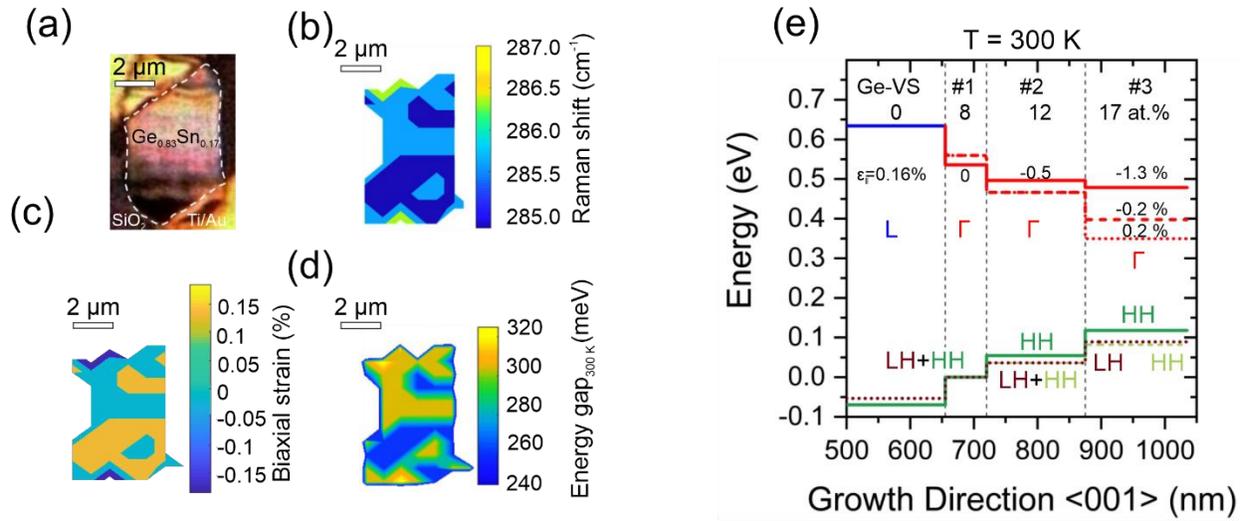

**Figure 3.** (a) Optical micrograph of a Ge$_{0.83}$Sn$_{0.17}$ membrane processed device; (b) The corresponding Ge-Ge Raman map; (c) Strain map in the membrane device; (d) Band gap energy variation in the processed membrane device; (e) Calculated 8×8 *k·p* band lineup (at 300 K) for the Ge$_{0.83}$Sn$_{0.17}$ at an in-plane biaxial strain $\varepsilon_\parallel = -1.3\ \%$ (as-grown) and $\varepsilon_\parallel = \pm 0.2\ \%$ (membrane).

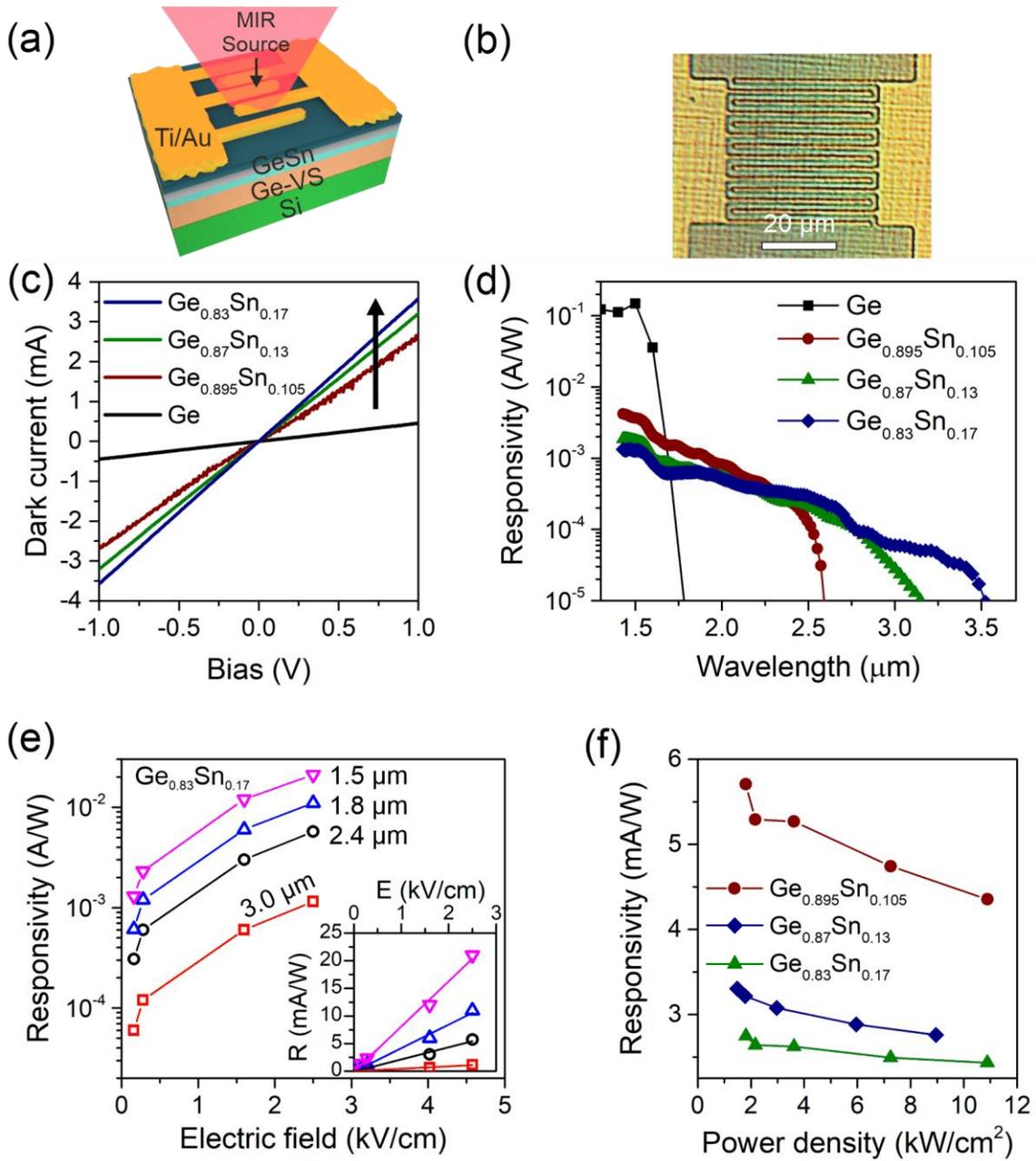

**Figure 4.** (a) Schematic illustration of the PD made of as-grown epitaxial layers; (b) Optical micrograph of the interdigitated photodetector device; (c) Dark current as a function of the bias voltage; (d) Spectral responsivity recorded for Ge and GeSn PDs; (e) Responsivity as a function of the electric field at different excitation wavelengths in $Ge_{0.83}Sn_{0.17}$ PD. Inset: the same but on linear scale; (f) Responsivity as a function of the excitation power density for GeSn PDs.

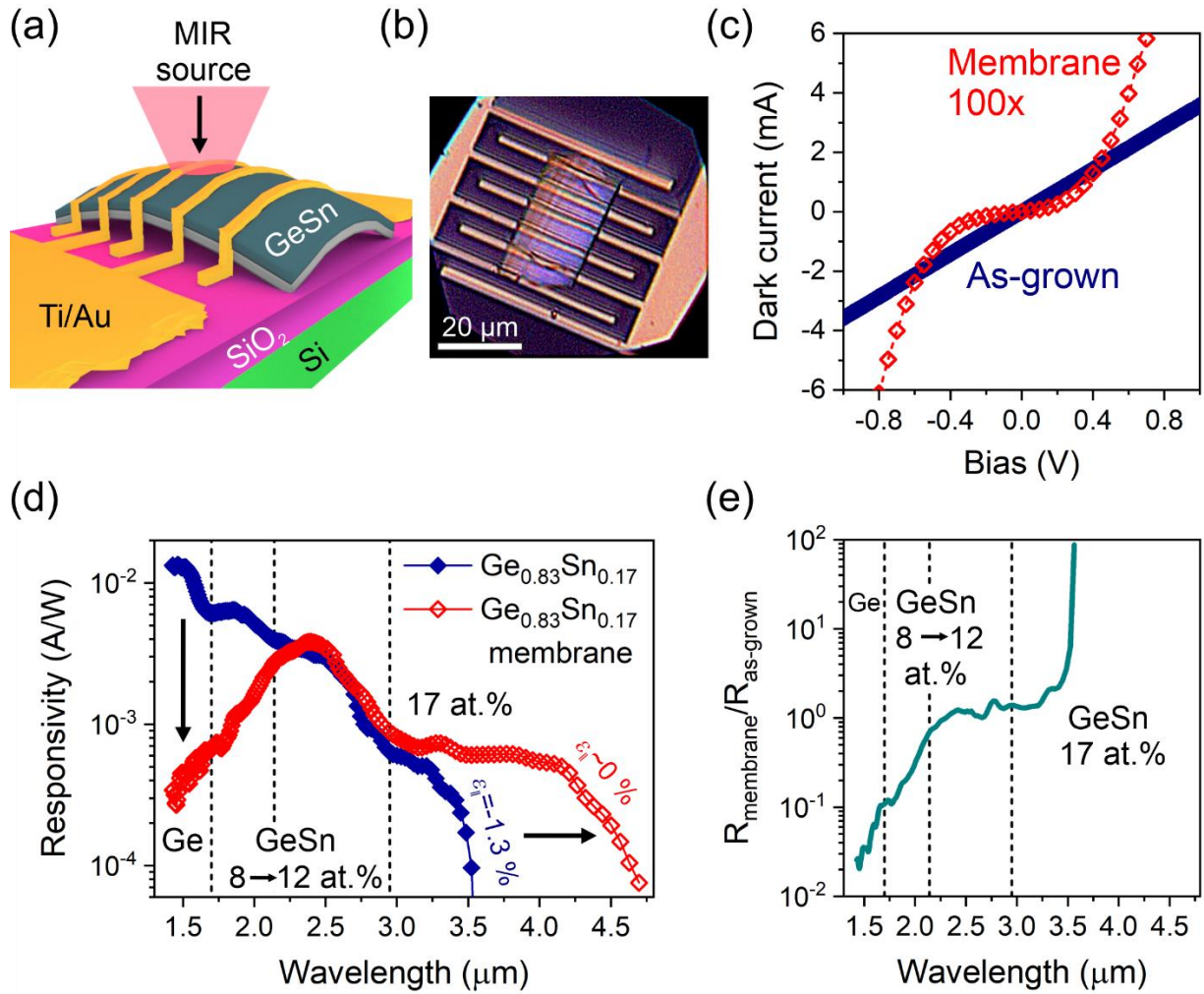

**Figure 5.** (a) Schematic of the transferred membrane PD; (b) Optical micrograph of a membrane PD on SiO$_2$/Si; (c) Dark current as a function of the bias voltage; (d) Responsivity as a function of the wavelength for the as-grown and membrane Ge$_{0.83}$Sn$_{0.17}$ devices; (e) Ratio of the responsivity of the Ge$_{0.83}$Sn$_{0.17}$ membrane to that of as-grown Ge$_{0.83}$Sn$_{0.17}$ PD as a function of the wavelength.

**TOC figure.**

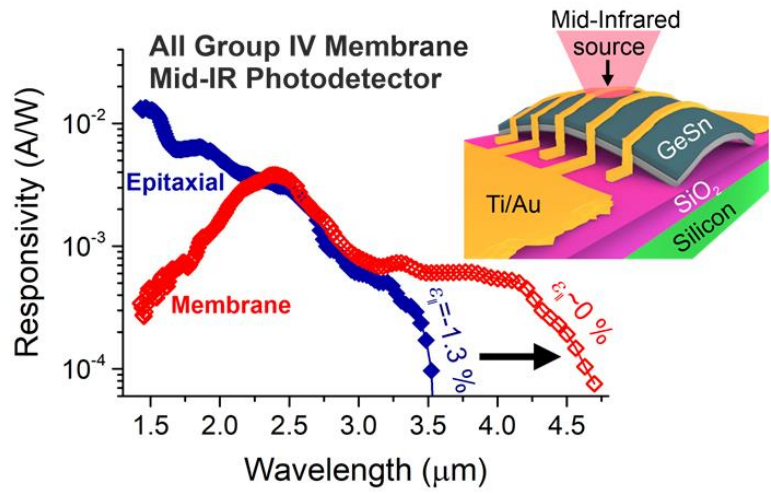

# Supporting Information

## All-group IV membrane for mid-infrared room-temperature photodetectors

Mahmoud R. M. Atalla,[1,ǂ] Simone Assali,[1,ǂ] Cédric Lemieux-Leduc,[1] Anis Attiaoui,[1] Aashish Kumar,[1] Salim Abdi,[1] and Oussama Moutanabbir[1,*]

[1]*Department of Engineering Physics, École Polytechnique de Montréal, C. P. 6079, Succ. Centre-Ville, Montréal, Québec H3C 3A7, Canada*

**Contents**

## S1. Structural parameters of the as-grown samples.

| Wafer | layers | thickness | Top layer strain |
|---|---|---|---|
| $Ge_{0.83}Sn_{0.17}$ | Ge-VS/$Ge_{0.93}Sn_{0.08}$/$Ge_{0.88}Sn_{0.12}$/$Ge_{0.83}Sn_{0.17}$ | 0.6 μm/0.065 μm/0.155 μm/0.16 μm | -1.3% |
| $Ge_{0.87}Sn_{0.13}$ | Ge-VS/$Ge_{0.95}Sn_{0.05}$/$Ge_{0.94}Sn_{0.06}$/$Ge_{0.91}Sn_{0.09}$/$Ge_{0.89}Sn_{0.11}$/$Ge_{0.87}Sn_{0.13}$ | 1.6 μm/0.24 μm/0.24 μm/0.17 μm/0.21 μm/0.12 μm | -0.52% |
| $Ge_{0.895}Sn_{0.105}$ | Ge-VS/$Ge_{0.95}Sn_{0.05}$/$Ge_{0.94}Sn_{0.06}$/$Ge_{0.92}Sn_{0.08}$/$Ge_{0.895}Sn_{0.105}$ | 1.6 μm/0.24 μm/0.24 μm/0.17 μm/0.15 μm | -0.38% |
| Ge-VS | Ge-VS | 1.6 μm | 0.16% |

**Table S1.** List of thickness, composition, and strain (in the upmost layer) for all samples investigated in this work.

## S2. Raman spectroscopy analysis.

Raman measurements were performed on the as-grown and membrane $Ge_{0.83}Sn_{0.17}$ samples to evaluate the residual strain in the TL. The Raman spectra are plotted in Fig. S1. Note that in the Raman spectra the contribution from the lower Sn content layers (#1-2) is negligible because of the penetration depth of the 633 nm excitation laser being limited to less than 30 nm, thus significantly smaller than the TL thickness of 160 nm. Among the recorded characteristic vibrational modes the Ge-Ge LO peak is measured at 292.5 ± 1.3 $cm^{-1}$ for as-grown sample, but it shifts to 287 ± 1 $cm^{-1}$ at the edge of the partially released membrane. The mode shifts further to 284.8 ± 1.3 $cm^{-1}$ for completely released membrane with the Ge VS and BL etched away. According to the Raman spectra in Fig. S1, also the Ge-Sn LO peak shifts from 255 $cm^{-1}$ to 246 $cm^{-1}$ upon strain relaxation in the released membrane. It is also noticed that the full width half maximum and the asymmetry have not changed as the strain is relaxed. According to our recent systematic studies [1,2] decoupling Sn content and degree of strain using Raman spectra analysis, the strain relaxation in the GeSn membrane could be calculated knowing the Sn content from the RSM map in Fig. 1a. The Ge-Ge LO peak position is linearly dependent on the strain and Sn content as

$$\omega_{Ge-Ge} = \omega_{0,Ge-Ge} + a\,y + b\,\varepsilon, \qquad (1)$$

where $\omega_{Ge-Ge}$ is the Ge-Ge Raman peak position of the strain relaxed membrane, $y$ is the Sn content, $\varepsilon$ is the strain, and $\omega_{0,Ge-Ge}, a, b$ are constants equal to 300.4 cm$^{-1}$, - 84 cm$^{-1}$, and - 491 cm$^{-1}$, respectively. Accordingly, the Ge-Ge LO peak shifts from 292.5 ± 1.3 cm$^{-1}$ to 284.8 ± 1.3 cm$^{-1}$, the strain changes from -1.3% to +0.2%, respectively, which indicates significant strain relaxation and even change from compressive to tensile strain.

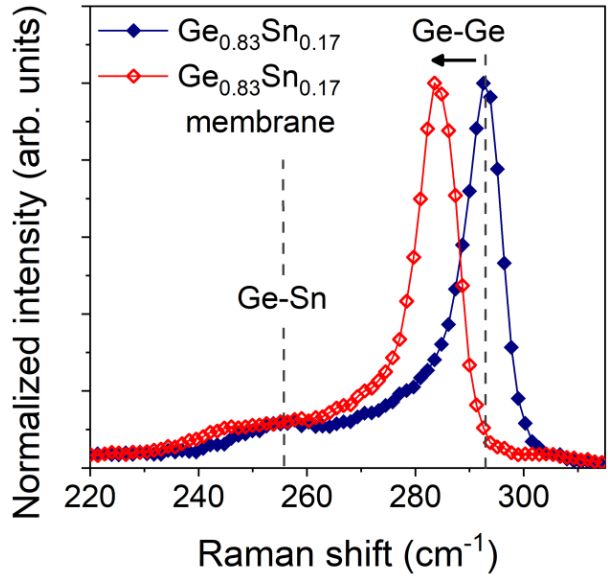

**Figure S1.** Raman spectra of the Ge$_{0.83}$Sn$_{0.17}$ as-grown and membrane samples, and dashed lines mark the GeGe and Ge-Sn peaks for as-grown sample.

## S3. Schematic of the photocurrent measurement setup.

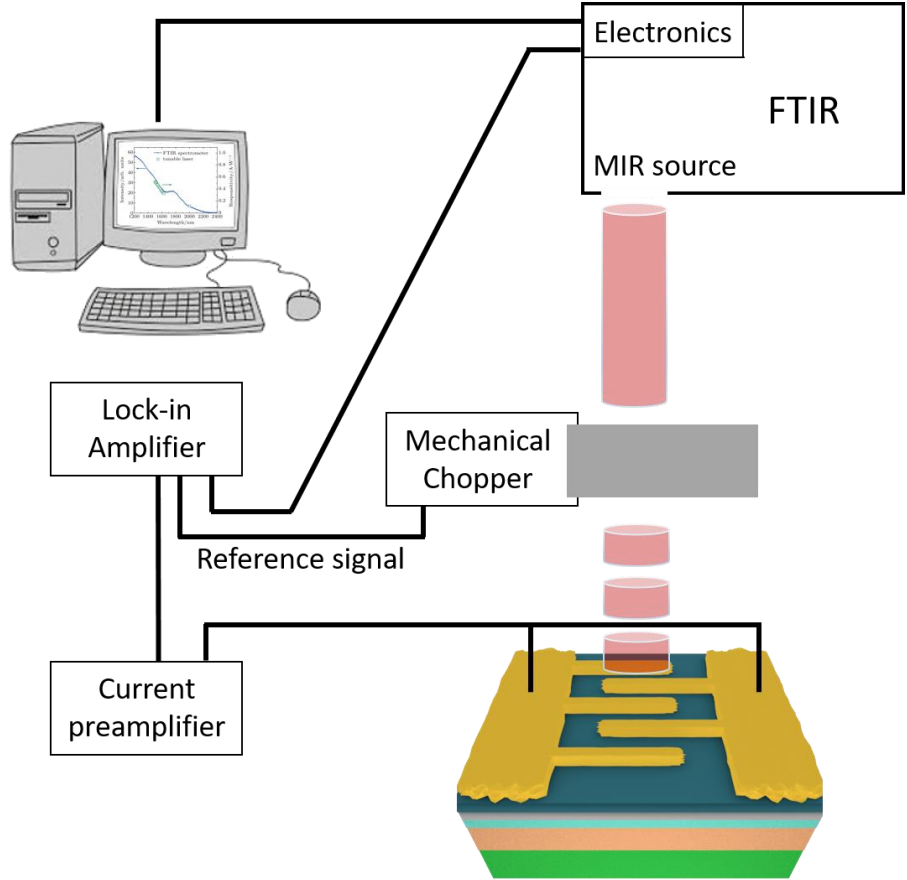

**Figure S2.** Schematic of the photocurrent measurement setup used to measure the Ge(Sn) PDs.

The MIR light source of the FTIR was incident on the GeSn device and the electrical signal was measured using a Zurich Instruments lock-in amplifier that was locked to the frequency of a chopper in the light path of the MIR light source, as shown in Fig. S2. Additionally, the photocurrent was measured at 1.55 μm wavelength using a tunable laser. The optical power was measured using a power meter and divided by the spot size area to estimate the optical power density. Furthermore, the effective electric field was calculated as the applied bias divided by the contact separation.

## S4. TLM parameters.

The contact resistance of Ti/Au device contacts and the sheet resistance of Ge(Sn) were estimated from transfer length measurements (TLM).[3] To this extent, Ti/Au contacts with sizes of 60×30 µm² and spacing of 2 µm were deposited on Ge-VS, GeSn samples (inset in Fig. 1f). From the linear dependence of the sheet resistance $R_{sh}$ as a function of the spacing (Fig. 3f), the specific resistivity $\rho_c$ was extracted for all samples, as listed in Table S2. The lowest resistivity $\rho_c = 1.2 \times 10^{-4}\ \Omega cm^2$ is achieved in Ge$_{0.83}$Sn$_{0.17}$, and increases to $3.0 \times 10^{-4}\ \Omega cm^2$ and in $3.9 \times 10^{-4}\ \Omega cm^2$ in Ge$_{0.87}$Sn$_{0.13}$ and Ge$_{0.895}$Sn$_{0.105}$, respectively. For Ge-VS, we obtained $\rho_c = 2.8 \times 10^{-3}\ \Omega cm^2$. Similar behavior is also observed for the sheet resistance $R_{sh}$. By maximizing (minimizing) $R_{sh}$ ($\rho_c$) the performance of a PD would improve, which is in agreement with the higher efficiency observed for Ge-VS as compared to GeSn.

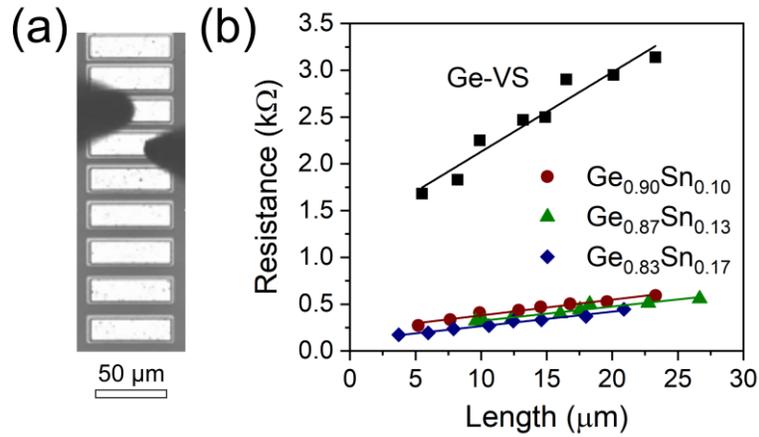

**Figure S3.** (a) Optical micrograph of the TLM device. (b) Resistance as a function of the length in TLM measurements.

| Sample | $L_T$ (µm) | $R_C$ (Ω) | $\rho_c$ (Ω.cm²) | $R_{sh}$ (Ω/sq) |
|---|---|---|---|---|
| Ge$_{0.83}$Sn$_{0.17}$ | 3.6 | 57.5 | 1.16 X 10$^{-4}$ | 888.72 |
| Ge$_{0.87}$Sn$_{0.13}$ | 5.7 | 92 | 2.94 X 10$^{-4}$ | 898 |
| Ge$_{0.895}$Sn$_{0.105}$ | 6.3 | 110 | 3.88 X 10$^{-4}$ | 974 |
| Ge VS | 7.5 | 673 | 2.82 X 10$^{-3}$ | 4990 |

**Table S2.** Parameters extracted from the TLM fit.

## S5. Capacitance-voltage measurements

Capacitance-Voltage (CV) measurements were carried out to extract their active carrier concentrations using Back-to-Back (B2B) Metal-Oxide-Semiconductor capacitors (MOScaps). The measurements were carried out in a Keithley 4200A-SCS parameter analyzer and at room temperature, and the chosen frequency for reliable dopant level extraction in GeSn is 1 MHz. The dielectric oxide used was e-beam deposited $SiO_2$ with a relative permittivity of 3.9, for this reason, a thickness of 23 nm was chosen. Since B2B MOScaps do not contain a metal-semiconductor interface, establishing ohmic contacts is not an issue in their fabrication and characterization. Finally, the analysis of the resulting CV curves was established through the following equation:

$$N_{sub} = \frac{2}{q\varepsilon_s A^2 \left(\frac{\Delta 1/c^2}{\Delta V_g}\right)}$$

Where $\varepsilon_s$ is the relative permittivity of the semiconductor in F.cm$^{-1}$, A is the area of the gate, and $\left(\frac{\Delta 1/c^2}{\Delta V_g}\right)$ is the averaged slope of the linear part in transition region of the CV curves. As for the values, the relative permittivity of GeSn was assumed to be close to pure Ge with a value of $\varepsilon_s = 16$; and a correction factor of 0.8 was used for the area in the B2B devices based on the dimensions of the ring geometry of the capacitor.

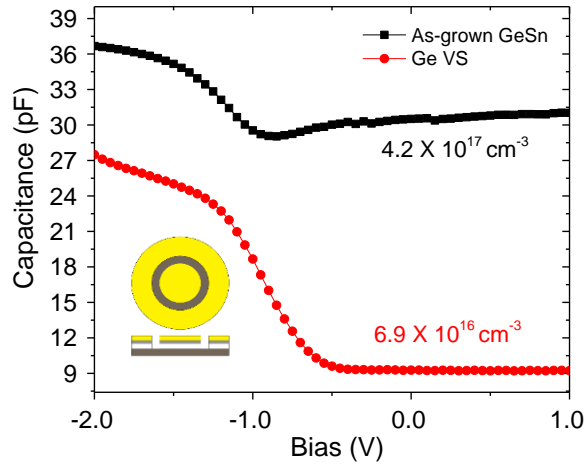

**Figure S4.** Comparison of the capacitance as function of voltage of the Ge VS and Ge$_{0.83}$Sn$_{0.17}$ samples.